\documentclass{jfm}
\usepackage{graphicx}
\usepackage{epstopdf, epsfig}
\usepackage{amsmath}
\usepackage{hyperref}
\usepackage{xcolor}

\shorttitle{SSM reduction of an inverted flag}
\shortauthor{Z. Xu, B. Kaszás, M. Cenedese, G. Berti, F. Coletti and G. Haller }

\title{Data-driven modeling of the regular and chaotic dynamics of an inverted flag from experiments}

\author{Zhenwei Xu\aff{1}, Bálint Kaszás\aff{1},
  Mattia Cenedese\aff{1}, Giovanni Berti\aff{2}
  Filippo Coletti\aff{1} \and
  George Haller\corresp{\email{georgehaller@ethz.ch}}\aff{1}
}

\affiliation{\aff{1}Department of Mechanical Engineering, ETH Zurich,
Switzerland
\aff{2}Department of Mechanical Engineering, Politecnico di Milano,
Italy
}

\begin{document}

\maketitle

\begin{abstract}
We use video footage of a water tunnel experiment to construct a 2D reduced-order model of the flapping dynamics of an inverted flag in uniform flow. The model is obtained as the reduced dynamics on a 2D attracting spectral submanifold (SSM) that emanates from the two slowest modes of the unstable fixed point of the flag. Beyond an unstable fixed point and a limit cycle expected from observations, our SSM-reduced model also confirms the existence of two unstable fixed points for the flag, which were found by previous {studies}. Importantly, the model correctly reconstructs the dynamics from a small number of general trajectories and no further information on the system. {In the chaotic flapping regime, we construct a 4D SSM-reduced model that captures the system's chaotic attractor.}
\end{abstract}


\section{Introduction}
Accurate modeling of fluid flows has consistently posed challenges due to the expensive numerical simulations required. The objective of reduced-order modeling is to decrease these costs while preserving accuracy and physical relevance. Methods like the proper orthogonal decomposition (POD) \citep{holmes} identify key modes from data, onto which the governing equations can be projected. Dynamic mode decomposition (DMD) \citep{Schmid2022Review} directly finds a linear system governing the time evolution. Although such approaches and their variants are powerful and widely used, the linear models they return cannot capture essentially nonlinear behavior, emphasizing the need for nonlinear reduced-order models, as noted by \cite{PageKerswell2019,LinotGraham2020,KaszasHaller2024}. Our focus here is on fluid-structure interactions, crucial for vegetation \citep{vegetation} as well as for wind energy production \citep{windturbine}. Exhibiting  strongly nonlinear behavior, this problem also requires nonlinear reduced-order models. Pioneering research on the construction of such models has been discussed in the review of \cite{dowell2001review}.

The dynamics of a conventional flag with a clamped leading edge in uniform flow, a classic fluid-structure interaction system, have been widely studied (see the review paper by \cite{flagReview}). The inverted variant of the same setup, with the trailing edge clamped has also gained significant interest recently. Figure \ref{fig:schematicview} illustrates the latter configuration for the inverted flag.


\cite{kim2013flapping} experimentally showed that in both air and water flow, the amplitude dynamics of an inverted flag exhibit three main regimes: a stable undeformed regime, a large-amplitude oscillatory regime (flapping), and a fully deflected regime. These dynamics depend on the Reynolds number, the {flag-to-fluid} mass ratio $\mu_{\rho}$ and the dimensionless bending stiffness $K_B$. These parameters are defined as 
\begin{equation}
    Re = \frac{\rho_f U L}{\mu}, \qquad\mu_{\rho} = \frac{\rho_s h}{\rho_f L}, \qquad K_B = \frac{B}{\rho_f U^2 L^3},
\end{equation}
where $\rho_f$ and $\rho_s$ are the densities of the fluid and the elastic sheet the flag is made of; $U$ is the free stream velocity; $H$ and $L$ are the height and the length of the sheet, respectively; $h$ is the thickness of the sheet; $\mu$ is the viscosity of the fluid; and $B=E h^{3} / 12\left(1-\nu^{2}\right)$ is the bending rigidity of the sheet with $E$ and $\nu$ denoting Young's modulus and Poisson's ratio. {The non-dimensional stiffness $K_B$ quantifies the ratio of elastic restoring forces in the flag to fluid dynamic pressure due to the impinging flow, while the mass ratio $\mu_{\rho}$ measures the relative significance of flag inertia compared to fluid inertia}.

In addition to the three dynamical regimes already mentioned, \cite{ryu2015flapping} numerically identified an intermediate regime, the deformed equilibrium, between the flapping mode and the undeformed equilibrium, which was confirmed by \cite{Yu2017} in experiments. \cite{gurugubelli2015} demonstrated that an inverted flag can generate significantly more strain energy than a conventional flag. Motivated by this finding, \cite{shoele2016} explored the application of an inverted piezoelectric flag in energy harvesting. 
\cite{sader2016large} conducted experiments at Reynolds number $Re=\mathcal{O} (10^4-10^5)$ and found that flapping is predominantly periodic but transitions to chaos as the flow speed increases. {\cite{sader2016large} also predicted and experimentally verified two unstable deflected equilibria for the flag in addition to its trivial unstable equilibrium at its undeflected state. The authors identified the mechanism for large-amplitude flapping using both theory and experiment.}

\cite{goza2018global} characterized the chaotic flapping regime from simulations for a much lower Reynolds number $Re=\mathcal{O} (10^2)$. Chaotic flapping develops when the stiffness is smaller than for the large-amplitude flapping regime, but not yet small enough for the totally deflected regime. They also conducted a stability analysis of the fully coupled equations and found that as $K_B$ decreases, the deformed equilibrium state undergoes a supercritical Hopf bifurcation to small-deflection deformed flapping. As $K_B$ decreases further, large amplitude flapping sets in.

While previous studies have offered great insight by analyzing extensive experimental and numerical data, it is desirable to model the behavior of such a nonlinear system including its various states and the transitions between them based on limited observations. Reduced-order modeling approaches have not yet been explored for the inverted flag setup. We therefore use the recently developed theory of spectral submanifolds (SSMs) to obtain a 2D, data-driven model for the inverted flag directly from experimental data. SSMs are continuations of a spectral subspace of the linearized system near a stationary state such as a ﬁxed point, a periodic orbit, or a quasiperiodic torus \citep{haller2016nonlinear}. Initially, only the smoothest {(or primary)} SSMs of stable stationary states and their reduced dynamics were constructed via Taylor-expansions. More recently, \cite{mixedModeSSM} extended the theory to include secondary SSMs of limited smoothness (fractional SSMs) as well as SSMs tangent to eigenmodes of mixed stability type (mixed mode SSMs). We will {approximate} the dynamics on an attracting, mixed-mode, {primary} SSM as a reduced-order model for the inverted flag.

In the large-amplitude flapping regime there are three unstable fixed points together with the limit cycle motion. They are all coexisting stationary states of the system. Under nonresonance conditions, SSM theory establishes the existence and uniqueness of SSMs attached to the middle equilibrium point. We show that the two deflected unstable equilibria and the stable limit cycle predicted by previous {studies} are all contained in this mixed-mode SSM. This enables us to derive a simple 2D data driven-model that captures all stable and unstable steady states of the flag as well as transitions among them. {In the chaotic flapping regime, we construct a reduced model on a 4D mixed-mode SSM that captures the chaotic attractor. We believe this SSM to be the first experimentally constructed inertial manifold for a chaotic attractor, obtained from the same methodology derived and illustrated by \cite{liu_2023} on numerical data. On a broader note, this study shows that SSM-reduced models can reliably identify not only stable fixed points, but unstable fixed points, limit cycles, heteroclinic orbit and even chaotic attractors using a moderate numbers of experimental observations.}

\section{Experimental setup and nonlinear model reduction method}\label{sec:exp_setup}
\subsection{Experimental apparatus and procedure}

\begin{figure}
    \centering
    \includegraphics[width=0.75\textwidth]{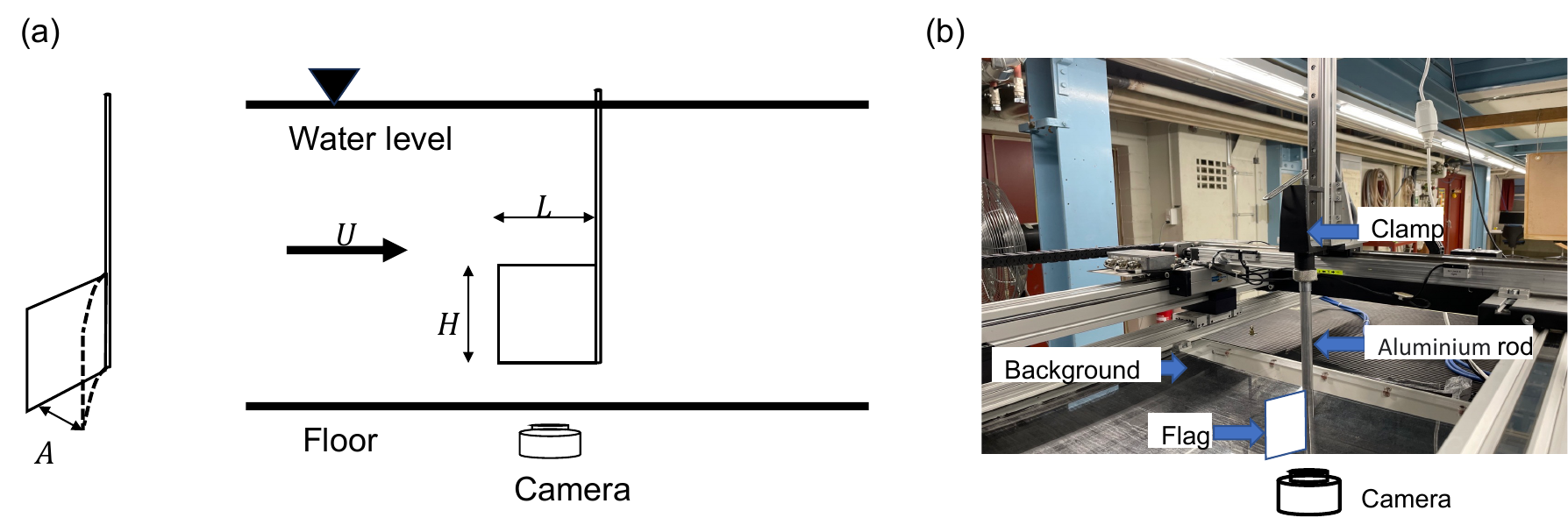}
    \caption{(a) Schematic view of the system with the water tunnel, linear actuator and the camera system. {We denote the flag's length by $L$, height by $H$ and thickness by $h$.} {The tip's deflection is denoted by $A$, defined as the distance between the tip of the flag in its deflected and undeflected states}. Since the flapping is symmetric, we choose one side to be positive and other side to be negative in sign. (b) Lab setup illustration.}
    \label{fig:schematicview}
\end{figure}

We carry out our experiments in a recirculating open channel facility. A contraction ensures laminar flow entering the $2.4 \mathrm{~m}$ long, $0.45 \mathrm{~m}$ wide test section, filled with water to a depth of $0.41 \mathrm{~m}$. {The tunnel has a smooth entry with a large contraction ratio designed to provide uniform and laminar flow. Separate PIV measurements indicate that the level of spatial non-uniformity and the root-mean-square temporal fluctuations are of the order of the measurement error, and therefore are not deemed significant for this study.} We conduct experiments using two flags of different bending rigidity in different ranges of flow velocities, as summarized in table \ref{tab:flagProp}. These yield two Reynolds numbers, $Re = 6\times 10^4$ and $Re = 10^5$, defined at the velocity at which the large-amplitude periodic flapping have been modeled. The flags consist of polycarbonate sheets (Young’s modulus $E = 2.38\times 10^9 ~Pa$, Poisson ratio $\nu = 0.38$, density $\rho = 1.2\times 10^3 ~kg\cdot m^{-3}$) clamped on an aluminum beam, 16 mm in diameter and rigidly mounted to the body of the channel. A schematic of the setup is shown in Fig. \ref{fig:schematicview}.
The motion of the bottom tip of the flag, marked with a small color strip, is captured by a digital camera (4 Megapixels, 120 fps) through the transparent floor, a black cover on the test section serving as background. Beside the deflection $A$, we evaluate the velocity $\dot{A}$ after applying a Gaussian filter to the position data. The uncertainty on the latter, due to finite imaging resolution and small bending of the beam, is estimated as 1\% of the total length of the flag.
We conducted two types of experiments to characterize both the steady-state and the transient dynamics. For the former, we analyze the behavior after initial transients have dissipated, while the latter involve releasing the flag from various deflected positions with zero initial velocity.

\begin{table}
  \begin{center}
\def~{\hphantom{0}}
  \begin{tabular}{lccc}
      parameter  & $Re=10^5$   &   $Re=6\times 10^4$ \\[3pt]
       $L$   & $150 \ mm$ & $150\ mm$ \\
       $H$   & $150\ mm$ & $150\ mm$ \\
       $h$  & $1.5\ mm$ & $0.8\ mm$ \\
       $U$   & $1.0\ m/s$ & $0.41\ m/s$\\
       $\mu_{\rho}$   & $0.012$ & $0.0064$\\
  \end{tabular}
  \caption{Flag properties in the two experiments; We used the same material, polycarbonate, for all experiments such that Young’s modulus, density and Poisson’s ratio remain the same. $U$ is the free stream velocity at which the periodic flapping behavior has been modeled.}
  \label{tab:flagProp}
  \end{center}
\end{table}

\subsection{Spectral submanifolds (SSMs) and data-driven reduced modeling of the flag dynamics}
The system under study is infinite-dimensional, governed by the coupled Navier-Stokes and Euler-Bernoulli partial differential equations (PDEs) for the fluid and solid, respectively. We discretize this set of PDEs to obtain a system of ordinary differential equations (ODEs) in the form: 
\begin{equation}\label{eq:firstorderSystem}
\dot{\boldsymbol{x}} = \mathsfbi{A} \boldsymbol{x}+\boldsymbol{f}_{nl}(\boldsymbol{x}), \quad \boldsymbol{x}\in\mathbb{R}^n,\quad\boldsymbol{f}_{nl}=\mathcal{O}(\left | \boldsymbol{x} \right |^2) \in C^{\infty}.
\end{equation}
The linearization of this system around its $\boldsymbol{x}=\boldsymbol{0}$ equilibrium is governed by the operator $\boldsymbol{A}$. Let $\lambda_1,...,\lambda_n \in \mathbb{C}$ denote the eigenvalues of $\boldsymbol{A}$. If these eigenvalues satisfy the nonresonance condition
\begin{equation}\label{eq:nonresonance}
    \sum_{j=1}^{n} m_j \lambda_j \neq \lambda_k, \quad m_j\in \mathbb{N},
\end{equation}
for $k=1,...,n$ and for all non-negative integers $m_j$ with $\sum_{j=1}^{n} m_j \geq 2 $, then system \eqref{eq:firstorderSystem} can be exactly linearized via a $C^{\infty}$ change of coordinates near $\boldsymbol{x}=\boldsymbol{0}$ (see \cite{Sternberg1958}). This implies that any spectral subspace $\mathcal{E}$ (i.e., any span of generalized eigenspaces) of $\boldsymbol{A}$ gives rise to a $C^{\infty}$ {spectral} submanifold (SSM) $\mathcal{W}(\mathcal{E})$ for the nonlinear system \eqref{eq:firstorderSystem} near the origin \citep{mixedModeSSM}. $\mathcal{E}$ is called a like-mode spectral subspace if the signs of the real parts of all $\lambda_j$ inside $\mathcal{E}$ are the same, and $\mathcal{E}$ is called a mixed-mode spectral subspace otherwise. Similarly, the SSM $\mathcal{W}(\mathcal{E})$ is called a like-mode SSM if the associated $\mathcal{E}$ is like-mode spectral subspace, and $\mathcal{W}(\mathcal{E})$ is called a mixed-mode SSM otherwise.

This $\mathcal{W}(\mathcal{E})$ is an invariant manifold tangent to $\mathcal{E}$ at $\boldsymbol{x}=\boldsymbol{0}$ and has the same dimension as $\mathcal{E}$. If all eigenvalues of $\boldsymbol{A}$ outside $\mathcal{E}$ have negative real parts, then $\mathcal{W}(\mathcal{E})$ is an attracting invariant manifold to which all neighbouring trajectories converge exponentially fast. In that sense, the dynamics on $\mathcal{W}(\mathcal{E})$ serves as a reduced-order model with which all nearby trajectories synchronize exponentially fast.

Applications of SSM-based model reduction to mechanical systems have been discussed by \cite{Exactmodelreduction2017,breunung2018explicit,haller2016nonlinear,ponsioen2018automated,ponsioen2020model,JAIN2018195,computeinvariantmanifolds}. Recently, a data-driven approach to SSM-reduction has been developed and a corresponding MATLAB package has been released by \cite{Cenedese2022,cenedese2022data}. 
As noted by \cite{Cenedese2022}, if observations of the full state space variable $\boldsymbol{x}$ are not available, one can use delay-embedding of a smaller set of observables to reconstruct SSMs based on Takens's embedding theorem \citep{takens1981}. Under appropriate nondegeneracy conditions, this theorem guarantees that an $m$-dimensional invariant manifold of system \eqref{eq:firstorderSystem} can be smoothly embedded into a generic observable space of dimension $p \geq 2m+1$. 

In the case of the inverted flag, the tip displacement $A(t)$ can be used to define a delay-embedded observable vector $\mathbf{y}(t) = (A(t), A(t+\Delta t), ..., A(t+ (p-1)\Delta t)) \in \mathbb{R}^{p}$. 
The SSM $\mathcal{W}(\mathcal{E})$ we seek to reconstruct in the space of this observable is the slowest SSM of the unstable fixed point corresponding to the undeflected position of the flag. {The smallest dimension of such a manifold required to capture the periodic flapping of the flag is two. This 2D manifold is the slowest mixed-mode SSM and acts as an attractor for the dynamics. Capturing faster decaying transients and more complicated dynamics would require the construction of higher-dimensional SSMs that also exist as long as the non-resonance conditions hold. As we will see, there is no trace of faster modes or other prominent frequencies in the large-amplitude flapping regime. In contrast, in the chaotic regime discussed in section 3.4, we use a 4D mixed-mode SSM.} 

The 2D SSM {we construct first} is tangent to the spectral subspace of the linearized dynamics that is spanned by the eigenvectors of a positive and a negative real eigenvalue. By Takens's theorem, the dimension of the observable space should then be at least $p=5$, given that the dimension of the SSM is $m=2$.

To construct the parametrization of the 2D SSM $\mathcal{W}(\mathcal{E})$ in the observable space, we follow the procedure developed by \cite{Cenedese2022}. Let $\mathbf{W}_0\in \mathbb{R}^{2\times p}$ be the matrix composed of the \textit{a priori} unknown vectors spanning the tangent space $\mathcal{E}$ of $\mathcal{W}(\mathcal{E})$ at the origin. We approximate $\mathbf{W}_0$ from singular value decomposition of trajectory segments passing by the undeflected position of the flag. Specifically, $\mathbf{W}_0$ is chosen as the span of the two leading singular vectors of the trajectory data matrix constructed for the observable $\mathbf{y}(t)$ {after excluding initial transients that contain modes beyond those used in constructing the slow SSM}. Reduced coordinates on $\mathcal{E}$ can be defined as $\boldsymbol{\eta} =\mathbf{W}_0 \mathbf{y}$, i.e. the projection of the observable coordinates onto $\mathcal{E}$. We aim to parametrize the SSM $\mathcal{W}(\mathcal{E})$ as a polynomial function $\boldsymbol{y} = \mathbf{V}_0\boldsymbol{\eta}  + \mathbf{V}\boldsymbol{\eta}^{2:l}$, where the vector $\boldsymbol{\eta}^{2:l}$ contains all scalar monomials of the variables $\eta_1$ and $\eta_2$ from degree $2$ to $l$ and $\mathbf{W}_0\mathbf{V}_0=\mathbf{I}$. Delay-embedded slow SSMs are nearly flat near the origin as deduced in general by \cite{Cenedese2022}, {for a moderate number of sufficiently small delays and signal with moderate time-derivative.} Based on this observation, we set $\mathbf{V}=\mathbf{0}$. {However, in the chaotic regime, we will utilize a non-flat SSM parameterized with a third-order polynomial owing to the larger time-lag.}

We use polynomials up to $M$-th order to represent the dynamics on {the SSM} $\mathcal{W}(\mathcal{E})$ in reduced coordinates as
\begin{equation}\label{eq:RD}
\begin{gathered}
    \dot{\boldsymbol{\eta}} = \mathbf{R}_0 \boldsymbol{\eta} + \mathbf{R}\boldsymbol{\eta}^{2:M}, \quad\boldsymbol{\eta} = (\eta_1, \eta_2)^T \in \mathbb{R}^2.
\end{gathered}
\end{equation}

To learn $\mathbf{R}_0$ and $\mathbf{R}$ from the data set of $K$ experiments $\mathbf{y}_1, \ldots, \mathbf{y}_K$, we first project them to $\mathcal{W}(\mathcal{E})$ in a direction orthogonal to $\mathcal{E}$ to obtain $\boldsymbol{\eta}_1, \ldots, \boldsymbol{\eta}_K$. We obtain the time derivatives  $\dot{\boldsymbol{\eta}}$ via numerical differentiation and formulate a minimization problem to find the best fitting matrices $(\mathbf{R}_0^*, \mathbf{R}^*)$ in formula \eqref{eq:RD} using the cost function
\begin{equation}\label{eq:opt}
\begin{gathered}
\left(\mathbf{R}_0^*, \mathbf{R}^*\right)=\arg\min _{\mathbf{R}_0,\mathbf{R}} \sum_{j=1}^K\left\|\dot{\boldsymbol{\eta}}_j-\mathbf{R}_0\boldsymbol{\eta}_j- \mathbf{R}\boldsymbol{\eta}_j^{2:M}\right\|^2.
\end{gathered} 
\end{equation}

{Our methodology, dynamics-based machine learning, integrates data-driven techniques with rigorous results from dynamical systems theory (\cite{haller2022dynamics}). }Among all the dynamic regimes, large-amplitude flapping tends to generate the most interest as it results in large strain energy suitable for energy harvesting. Our objective is to find the slow SSM of the undeformed state with its reduced dynamics in order to model all coexisting stationary states (deformed equilibria on two sides, undeformed equilibrium in the middle and a stable limit cycle), as well as transition orbits among them. 

\section{Results}\label{sec:result}

\subsection{Dynamical regimes of an inverted flag}
Previous findings indicate that as the stiffness $K_B$ decreases, the flag experiences a sequence of different dynamic regimes, as discussed in the Introduction. Starting with an undeformed static equilibrium, it progresses through small deflections, small-amplitude flapping, large-amplitude flapping, chaotic flapping, and finally, fully deflected state with the free end pointing downstream as a conventional flag. 
\cite{kim2013flapping} and \cite{Yu2017} report qualitatively similar bifurcation diagrams after comparing results from water and wind tunnels. In our experiments, we focus on mass ratio $\rho_s h/\rho_f L = \mathcal{O}(10^{-2})$ and aspect-ratio $H/L=1$.

\begin{figure}
    \centerline{\includegraphics[width=0.85\textwidth]{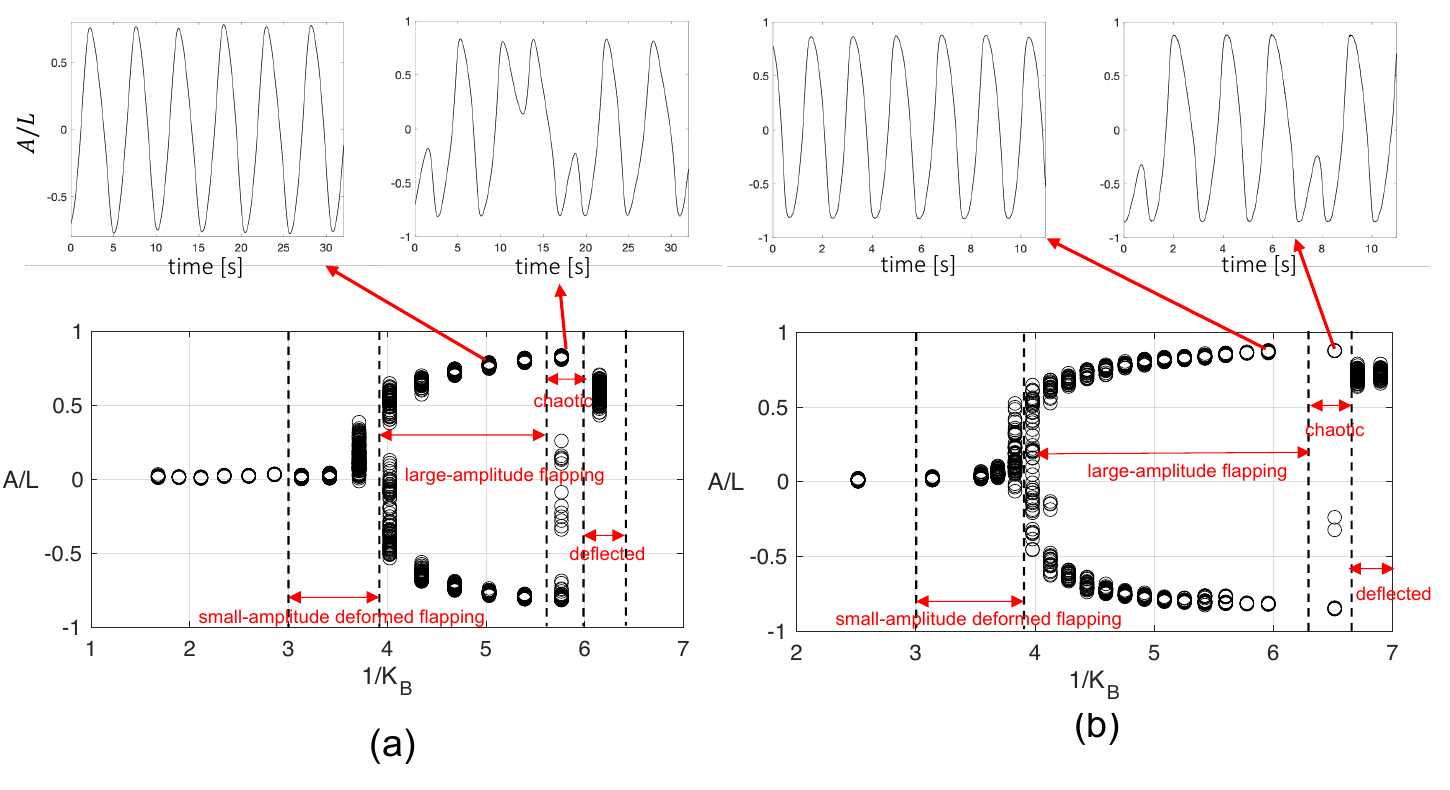}}
    \caption{Bifurcation diagrams of the inverted flag as a function of stiffness. Under decreasing stiffness $K_B$, the flag experiences {different dynamics}, as seen in the plots. {Also shown are representative time histories for the different regimes.} (a) $Re = 6\times 10^4$ and (b) $Re = 10^5$}\label{fig:bif}
\end{figure}

We plot the tip displacements against the stiffness $1/K_B$ to construct the bifurcation diagram for the flag {in Fig.\ref{fig:bif}}.  Following \cite{goza2018global} and \cite{TAVALLAEINEJAD2021}, we construct a Poincaré-section based on the tip velocity crossing zero. {Detecting the first bifurcation of static divergence is challenging due to imperfections in the mounting procedure and the initial curvature of the flag, as noted by \cite{TAVALLAEINEJAD2021}. In the small-amplitude deformed flapping regime, the flag oscillates around the deflected equilibrium with small amplitude. As the flag oscillates around the deflected equilibrium instead of the undeflected position, it exhibits asymmetry in the bifurcation diagram in Fig.\ref{fig:bif}. This observation is consistent with findings by \cite{TAVALLAEINEJAD2021} and \cite{goza2018global}}. 

As shown in Fig. \ref{fig:bif}, the bifurcation diagrams for both investigated $Re$ are quantitatively similar. This aligns with the findings of \cite{sader2016large}, that the flapping dynamics regimes are insensitive to changes in $Re$. We find that the first bifurcation leading to
large amplitude flapping occurs for $0.18\leq K_B \leq 0.25 $ at $Re = 6\times 10^4$ and for $0.16\leq K_B \leq 0.25 $ at $Re = 10^5$. This is consistent with the experiments by \cite{kim2013flapping} who observed such bifurcation for $0.2 \leq K_B \leq 0.25$ at a similar mass ratio. { This is expected, despite the variation in aspect ratios between our experiments and those conducted by \cite{kim2013flapping}. Our aspect ratio of 1, compared to the range of 1.3 to 2 in \cite{kim2013flapping}, is large. As indicated by \cite{sader2016large} and \cite{TAVALLAEINEJAD2021}, this similarity in aspect ratios results in a comparable range of bifurcation behaviors.}

We observe that under decreasing stiffness $K_B$, the inverted flag explores the same dynamic regimes that have been reported by others, as stated in the Introduction. Specifically, we note that the flag {starts from the straight configuration} in the stable undeformed equilibrium, then buckles with small deflections at a stable deflected equilibrium, progresses to small-amplitude asymmetric deformed flapping and then large-amplitude flapping. Chaotic behavior, {characterized by a positive Lyapunov exponent (as will be shown in Section 3.4)}, is also observed before the flag ultimately becomes fully deflected. 

\subsection{The reduced-order model constructed at $Re = 6\times 10^4$}\label{sec:SSM_Re1} 
\begin{figure}
    \centering
    \includegraphics[width=.9\textwidth]{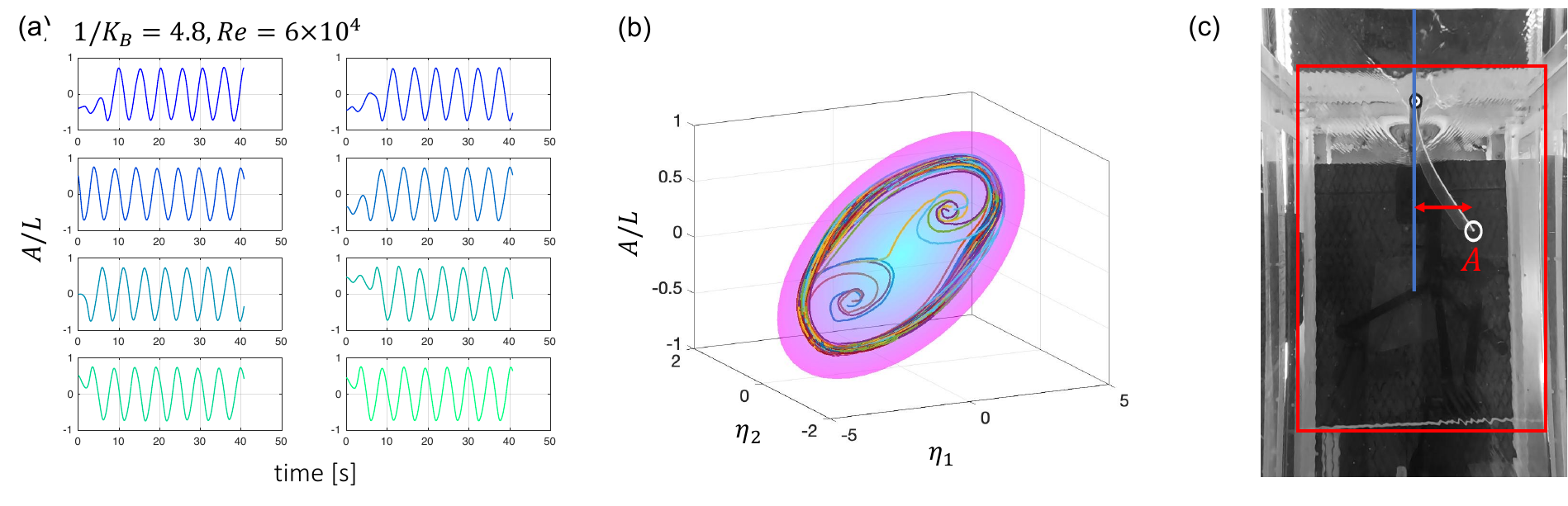}
    \caption{(a) Eight sample trajectories collected during the transient experiments. They show transient dynamics before settling down to the stable limit cycle oscillation. (b) Trajectories in the reduced coordinates and the SSM (pink) obtained using the training data. Here $\eta_1$ and $\eta_2$ are the reduced coordinates and the vertical axis denotes the amplitude of the tip deflection. (c) A snapshot of the experimental video. The blue line indicates the undeflected position of the flag; a white circle shows the position of the tip of the flag, the tip deflection is represented by the red arrow.}\label{fig:transient_traj}
\end{figure}

Having explored the dynamical regimes of the inverted flag, we now seek to find an SSM-reduced model in the large-amplitude flapping regime using the data-driven \textit{SSMLearn} methodology of \cite{Cenedese2022}. We obtain the necessary transient trajectories from experiments initialized with nonzero deflection and zero velocity at $Re = 6\times 10^4$. We apply delay-embedding to the tip displacement time-series to represent the trajectories in the observable space. The processed trajectories are shown in Fig. \ref{fig:transient_traj}a. To discover the dynamics in a larger subset of the observable space, we choose the initial deflections of the flag in the range of large-amplitude deflection, especially near the three fixed points.

The undeformed equilibrium is known to become an unstable saddle-type fixed point via a supercritical pitchfork bifurcation \citep{sader2016large} of the deformed equilibrium. {This instability arises from a divergence mechanism, in which the balance between aerodynamic forces and the elastic restoring force results in buckling.} {The} two slowest eigenvalues {of the undeformed equilibrium} are real {and of opposite sign}. The non-resonance condition \eqref{eq:nonresonance} {generically} holds for a {real-life dissipative physical system, such as ours}. Therefore, we {are justified to} fit a 2D mixed-mode SSM to the experimental data. We first divide the collected trajectories { after stripping them of their initial transients,} into a training set and a test set. The training set is used to construct $\mathcal{W}(\mathcal{E})$ from singular-value decomposition and its reduced dynamics \eqref{eq:RD} by solving the optimization problem \eqref{eq:opt}. The test set is reserved to evaluate the performance of the SSM-based model. In total, we collect 20 trajectories, 4 of which are reserved for testing and 16 are used for training.  In Fig. \ref{fig:transient_traj}a-b, we  show the training data we use together with the two-dimensional SSM $\mathcal{W}(\mathcal{E})$.

{For an embedding dimension $p$}, we extract the SSM $\mathcal{W}(\mathcal{E})$ from the training data and project the trajectories of the training set in the observable space on $\mathcal{W}(\mathcal{E})$. We then use the general polynomial form in Eq.\eqref{eq:RD} {with order $M$} to find the best fitting reduced dynamics. To determine these parameters for our non-regularized polynomial regression, {we test various combinations of $M$ and $p$. We then select the combination that yields the lowest reconstruction error as discussed in \cite{Cenedese2022}. Takens's theorem sets a minimum required embedding dimension, yet, a higher-dimensional embedding may improve reconstruction accuracy. We repeat this optimization for each Reynolds number considered here.} For the embedding dimension, we choose $p=25$  and for the polynomial order, we select $M=11$ in \eqref{eq:RD} after the {optimization at $Re=6\times 10^4$}.

\begin{figure}
    \centering
    \includegraphics[width=0.85\textwidth]{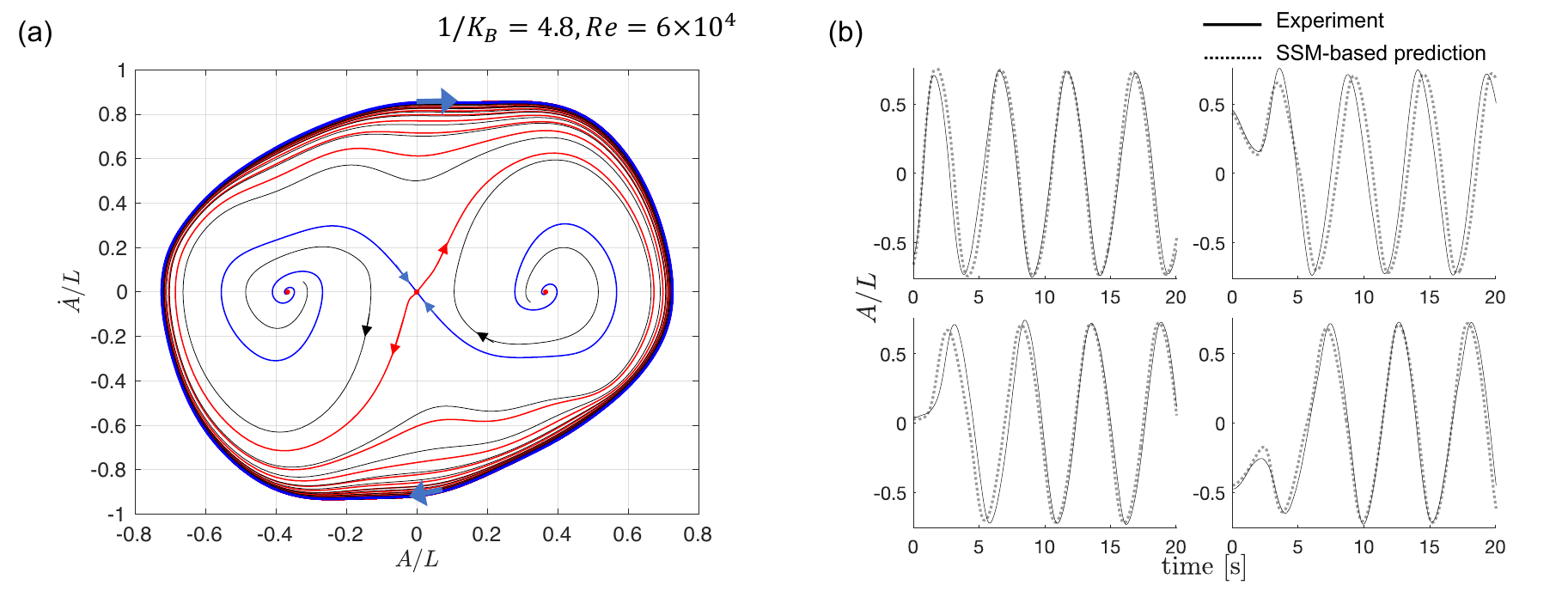}
    \caption{(a) The nonlinear phase portrait in the tip displacement and velocity coordinates at $Re=6\times10^4$. The red curve represents the unstable manifold, while the blue curve denotes the stable manifold of the undeformed state. The closed blue curve on the exterior shows the limit cycle, with two black curves depicting trajectories initialized near unstable deformed equilibria. Three fixed points are shown in red dots {and arrows indicate the trajectory orientations}.(b) Our predictions of the tip motion versus the test trajectories at $Re=6\times10^4$. The label "Experiment" refers to the experimentally measured test trajectories, while "Prediction" refers to predictions by the SSM-reduced model. }\label{fig:phyPhase}
\end{figure}

The reduced-order model \eqref{eq:RD} is a planar dynamical system. Inspection of the vector field defining the right-hand side of this model ODE reveals three coexisting unstable fixed points and a stable limit cycle. {We solve this ODE numerically to generate the trajectories shown in Fig. \ref{fig:phyPhase}a. The red points represent unstable fixed points, while the outer blue contour marks the limit cycle.} Physically, the fixed points capture two deformed unstable equilibira in symmetric positions and one undeformed equilibrium in the middle. {This agrees with the findings of \cite{sader2016large}}, and uncovers the {exact} geometry of transitions from these fixed points to the stable limit cycle describing the large-amplitude flapping of the flag. {In Fig.\ref{fig:phyPhase}b, the first 5 seconds, and in Fig.\ref{fig:highRePortrait}b, the first 2 seconds, illustrate the flag's transition from one unstable steady state to the attracting limit cycle. In Fig.\ref{fig:phyPhase}a and \ref{fig:highRePortrait}a, the blue curves connecting the fixed points represent heteroclinic transitions.} Note that we have not used any knowledge of the existence of the unstable deflected fixed points and the limit cycle: they were discovered from a model obtained using general trajectories. This shows the predictive power of our SSM-based reduced-order model.

Our low-dimensional model is also geometrically interpretable. The phase portrait of the planar system is influenced by the fixed points and their stable and unstable manifolds that we can find by time integration of the model. We initialize trajectories close to the three fixed points and advect them with the reduced model. We then transform the trajectories to the observable space and obtain the corresponding tip displacements as well as velocities. This results in a  phase portrait purely in terms of physical variables, as shown in Fig. \ref{fig:phyPhase}a. The heteroclinic orbits that connect the undeformed equilibrium with the deformed equilibria are shown in blue lines.

To examine the predictive power of our SSM-reduced model more closely, we reconstruct the test set. By running the SSM-reduced model on the same initial conditions numerically, we can advect the predicted trajectory, transform to the observable space and compare the predicted and the actual tip displacements. Figure \ref{fig:phyPhase}b shows the predictions of the model as well as the experimental data, which are in close agreement.

\subsection{The reduced-order model constructed at $Re = 10^5$}
In order to investigate the effect of the change in Reynolds number on the SSM-reduced model, we also perform the same experiments as in section \ref{sec:SSM_Re1} at $Re=10^5$. At this Reynolds number, a total of 17 transient trajectories are collected of which 13 are used for training and 4 for testing. 

Since the data set at this Reynolds number is intrinsically different from the one at lower Reynolds number, we perform again {an optimization} on the training data to determine the optimal values for $p$ and $M$. The dimension of the manifold remains two as the spectral subspace $\mathcal{E}$ has the same dimension and stability type. Based on a minimization of the trajectory reconstruction error, we chose the dimension of the observable space $p=9$ and the polynomial order $M=11$ in \eqref{eq:RD} after the {optimization}. 

\begin{figure}
  \centering
  \includegraphics[width=0.85\textwidth]{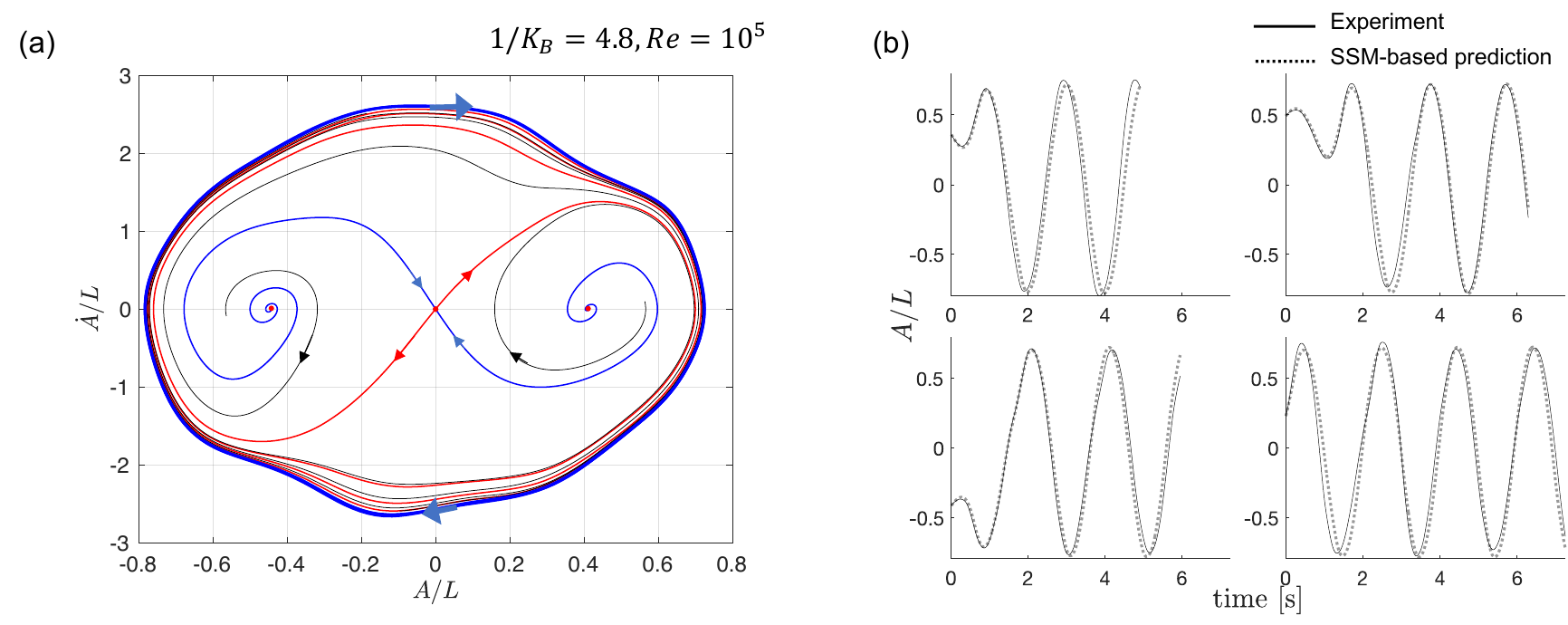}
  \caption{Same as Fig. \ref{fig:phyPhase} but for $Re=10^5$. }
  \label{fig:highRePortrait}
\end{figure}

We construct the phase portrait in the same way as before, by releasing initial conditions near the three fixed points and integrating the reduced model. Then we transform the trajectories back to the tip displacements and velocities shown in Fig. \ref{fig:highRePortrait}a. The SSM at higher Reynolds number also contains three unstable fixed points and one stable limit cycle. The reconstructed trajectories are obtained by integrating the model from the initial conditions of the test set. As shown in Fig. \ref{fig:highRePortrait}b, we again obtain accurate predictions over the test data set. Therefore the predictive power of our SSM-reduced model does not change as we double the Reynolds number.

\subsection{{Model reduction in the chaotic regime at $Re=6\times 10^4$}}
{The chaotic behavior bifurcates from the stable periodic flapping, as we further decrease $K_B$. We continue to seek the chaotic attractor embedded in a mixed-mode SSM, whose dimension is now necessarily larger than two. In other words, we seek an SSM that serves as an inertial manifold (\cite{foias_1988, LinotGraham2020}), as demonstrated by \cite{liu_2023}.}

{The dynamics of the flag close to the undeflected state resemble those of a buckling beam owing to their common divergence mechanism \citep{sader2016large}. We therefore expect that all eigenvalues, other than those associated to $\mathcal{E}$, are complex. As a result, the subspace $\mathcal{E}'$ of the second slowest linear decay rate is two-dimensional, making $\mathcal{W}(\mathcal{E}\oplus \mathcal{E}')$ a 4D mixed-mode SSM. Using the false nearest neighbors method \citep{kantz_book}, we find that the chaotic time series of the tip displacement, shown in Fig. \ref{fig:bif}, can indeed be embedded in four dimensions. }

{We collect a total of 11 trajectories in the chaotic regime at $Re=6\times 10^4$ and $1/K_B = 5.9$, and reserve one of them for testing. Based on the false nearest neighbors method, we select $4\Delta t$ as the time-lag for delay-embedding and embed the signal in a $p=9$-dimensional observable space. Due to this larger time-lag, the SSM can no longer be assumed flat. To this end, we parametrize it via an $l=3$-degree polynomial of the reduced coordinates $\boldsymbol{\eta}=(\eta_1, \eta_2, \eta_3, \eta_4)\in\mathbb{R}^4$.} 

{Since polynomial approximation for chaotic SSM-reduced dynamics was generally found inaccurate by \cite{liu_2023}, we use radial basis functions instead. Then the next-step prediction for the reduced coordinate $\boldsymbol{\eta}$ is written as }
\begin{equation}
    \boldsymbol{\eta}_{n+1} = \mathbf{F}(\boldsymbol{\eta}_n)=\sum_{i=1}^{K} \mathbf{C}_i k(||\boldsymbol{\eta}_n - \boldsymbol{\eta}_i||),
\end{equation}
{where $k$ is a radial kernel function depending on the distance of $\boldsymbol{\eta}_n$ from the points $\boldsymbol{\eta}_i$ in the training set. The coefficients $\mathbf{C}_i$ are determined by linear regression. We select a linear kernel by letting $k(r)=r$. Note that now the reduced dynamics is approximated as a discrete map instead of the ODE \eqref{eq:RD}, i.e., the time evolution of a trajectory is obtained by iteratively applying the map $\mathbf{F}$ starting from its initial condition.}

\begin{figure}
  \centering
  \includegraphics[width=1\textwidth]{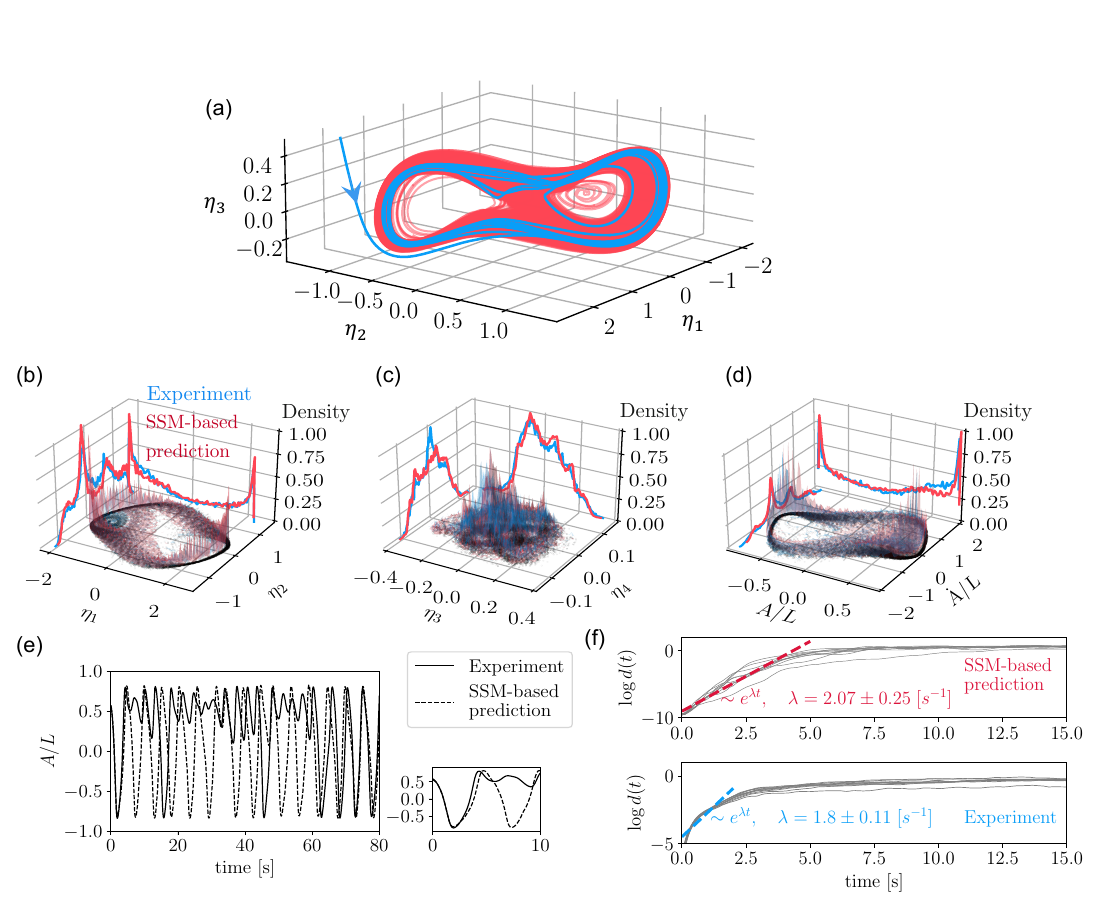}
  \caption{(a) Reconstructed chaotic attractor (red) in the reduced phase space. (b)-(d)  The invariant measure of the chaotic attractor determined from experiments (blue) and from the model (red). (e) Prediction of the test trajectory, with the initial 10 s magnified in the inset. (f) Logarithmic separation ($d(t)$) in the experimental data and in the model (see the main text). A linear fit to the initial section is computed for each grey curve corresponding to separate trajectories. The average exponent $\lambda$ and its standard deviation are reported. }
  \label{fig:chaotic_figure}
\end{figure}

{We demonstrate that the SSM-reduced model accurately captures the dynamical invariants of the system, such as its invariant measure, and leading Lyapunov exponent. In Fig. \ref{fig:chaotic_figure} we show the invariant measure of the chaotic attractor. We compute the empirical distribution from the experimental data, as well as from model-trajectories on the SSM. We find that the predicted distribution is reasonably close to the observed one.}

{We then estimate the leading Lyapunov exponent directly from the experimental data (\cite{kantz_book}). We select a number of reference points along the experimental trajectories and determine their nearest neighbors in the observable space. We measure the separation between the reference points and their nearest neighbors and compute the average-separation $d(t)$. The largest Lyapunov exponent $\lambda$ is then determined from the initial exponential trend of $d(t)$. This process is repeated for all trajectories.}

{To compute $\lambda$ from the 4D SSM-reduced model, we iterate the model starting from the same reference points as before and from a small (of size $10^{-4}$) perturbation to those points. The Lyapunov-exponents we have computed are reported in Fig. \ref{fig:chaotic_figure}f. We have estimated that the exponential increase in $d(t)$ is true for approximately $2.5-3$ s. Note that the actual value of $\lambda$ may be sensitive to this choice. Nevertheless, we find that $\lambda$ computed from the SSM-reduced model is in the order of the one estimated from the experiments.}

{In addition, we also show the prediction for the time evolution of the test trajectory in Fig. \ref{fig:chaotic_figure}e. As expected, due to sensitive dependence of trajectories on initial conditions, the prediction is only accurate for short times. We find, however, that this time is comparable to the Lyapunov-time, $1/\lambda.$}

\section{Conclusion}
In this paper, we have constructed a nonlinear reduced-order model for an inverted flag from experimental data collected in its large amplitude flapping regime. We have performed experiments to construct the canonical bifurcation diagram of the inverted flag that aligns with those reported in the literature. We have demonstrated that SSM-reduced nonlinear models can capture three coexisting unstable fixed points and a stable limit cycle, transitions {among them, and even a chaotic attractor. In the non-chaotic case,} we started with the undeformed unstable equilibrium as our anchor point and discover the other two unstable fixed points. Although these unstable fixed points were {reported by \cite{sader2016large}}, our reduced-order model did not utilize this information. Additionally, our SSM-reduced model has provided reliable predictions for transitions between those states, which have been unexplored in prior studies. In particular, we observed that initial conditions in the neighborhood of either one of the deformed states synchronize with the limit cycle dynamics, but their oscillations are in anti-phase, as evidenced by the black curves of Fig. \ref{fig:phyPhase}a and Fig. \ref{fig:highRePortrait}a. These types of orbits take the longest time to converge to the limit cycle. We have also found heteroclinic orbits connecting the undeformed equilibrium with the deformed states. 

{While previous studies \citep{goza2018global,TAVALLAEINEJAD2021} presented phase portraits determined by tip displacement and velocity, these plots alone do not yield a low-order model for predicting system behavior or fixed point locations. The power of the SSM approach lies in its ability to generate a robust and physically interpretable model that can also accommodate additional external forcing (see \cite{Cenedese2022}).} {By approximately doubling the Reynolds number, we have confirmed that the dynamics of the system are not sensitive to changes in Reynolds number in the considered regime $Re=\mathcal{O} (10^4-10^5)$. Our approach is fully data-driven, necessitating the provision of transient signals. In cases where the system is more complex and possesses additional modes, more trajectories originating from initial conditions associated with those modes are required}.

{In the chaotic flapping regime, we used the same approach to derive a 4D SSM-reduced model to capture the chaotic attractor of the system. One could also reconstruct the bifurcation diagram in Fig.\ref{fig:bif} from the model either by constructing independent models for each $K_B$, or by constructing models that depend on $K_B$ (\cite{KaszasCenedeseHaller2022}).}

Our 2D SSM-reduced models create an opportunity to design a closed-loop controller for the inverted flag. A possible actuation mechanism is the rotation of the trailing edge of the flag, as pursued by \cite{TangDowell2013} in a different setting. Possible control objectives range from the stabilization or elimination of select unstable equilibria through the reduction or enhancement of the limit cycle amplitude to {the suppression of the chaotic attractors}. Building on the recent SSM-reduced control methodology used in model-predictive control of soft robots \citep{alora2023data,alora2023practical}, the development of a closed-loop, SSM-reduced controllers for the inverted flag is currently underway.

\backsection[Supplementary material]{The code and data used in the paper is available in the repository \url{https://github.com/haller-group/SSMLearn}.}

\backsection[Funding]{This research was supported by the Swiss National Foundation (SNF) Grant No. 200021\_214908.}

\backsection[Declaration of interests]{The authors report no conflict of interest.}
\bibliographystyle{jfm}
\bibliography{jfm-instructions}

\end{document}